\documentclass{iaus}
\usepackage{graphicx}

\title[Chemical evolution of the Magellanic Clouds] 
{Chemical evolution of the Magellanic Clouds \\ based on planetary nebulae}

\author[Walter J. Maciel, Roberto D. D. Costa \& Thais E. P. Idiart]   
{Walter J. Maciel, Roberto D. D. Costa \and Thais E. P. Idiart}

\affiliation{Astronomy Department, University of S\~ao Paulo,\\ Rua do Mat\~ao 1226,
05508-900, S\~ao Paulo SP, Brazil \\ email: {\tt maciel@astro.iag.usp.br} 
\\ {\tt roberto@astro.iag.usp.br}, {\tt thais@astro.iag.usp.br}}

\pubyear{2008}
\volume{256}  
\pagerange{119--126}
\jname{The Magellanic System: Stars, Gas, and Galaxies}
\editors{Jacco Th. van Loon \& Joana M. Oliveira, eds.}
\begin{document}

\maketitle

\begin{abstract}
Planetary nebulae (PN) are an essential tool in the study of the chemical 
evolution of the Milky Way and galaxies of the Local Group, particularly the 
Magellanic Clouds. In this work, we present some recent results on the determination 
of chemical abundances from PN in the Large and Small Magellanic Clouds, and compare 
these results with data from our own Galaxy and other galaxies in the Local Group. 
As a result of our continuing long term program, we have a large database comprising 
about 300 objects for which reliable abundances of several elements from He to Ar 
have been obtained. Such data can be used to derive constraints to the nucleosynthesis 
processes in the progenitor stars in galaxies of different metallicities. We also 
investigate the time evolution of the oxygen abundances in the SMC by deriving the 
properties of the PN progenitor stars, which include their  masses and ages. We have 
then obtained an age-metallicity relation taking into account both oxygen and [Fe/H] 
abundances. We show that these results have an important consequence on the star 
formation rate of the SMC, in particular by suggesting a star formation  burst in 
the last 2-3 Gyr. 
\end{abstract}

\firstsection 
\section{Introduction}

Planetary nebulae (PN) are an essential tool in the study of the chemical evolution of 
the Milky Way and galaxies of the Local Group, particularly the Magellanic Clouds 
(see for example \cite[Maciel et al. 2006a]{maciel2006a}, \cite[Richer \& McCall 2006]
{richer}, \cite[Buzzoni et al. 2006] {buzzoni}, \cite[Ciardullo 2006)] {ciardullo}. 
As the offspring of stars within a reasonably large mass bracket 
(0.8 to about 8 solar masses), PN encompass an equally large age spread, as well 
as different spatial and kinematic distributions. They usually present bright emission 
lines and can be easily distinguished from other emission line objects, so that their 
chemical composition and spatiokinematical properties are relatively well determined. 
In this work, we present some recent results on the determination of chemical abundances 
from PN in the Large and Small Magellanic Clouds, and compare these results with data 
from our own Galaxy and other galaxies in the Local Group. We also investigate the time 
evolution of the oxygen abundances in the  SMC by deriving an age-metallicity relation 
for this object.

\section{Abundances}

Planetary nebulae and HII regions are particulary useful to study chemical abundances 
in the Galaxy as well as in other objects of the Local Group, especially since the advent 
of 4m class telescopes. While HII regions reflect the present chemical composition of the 
star-forming systems, PN can trace the time evolution of the abundances, especially when 
an effort is made to establish their age distribution (see for example Maciel et al. 2005). 
The elements S, Ar and Ne are probably not produced by the PN progenitor stars, as they are 
manufactured in the late evolutionary stages of massive stars. Therefore, S, Ar and Ne 
abundances as measured in PN should reflect the interstellar composition at the time the 
progenitor stars were formed. Since in the interstellar medium of a star-forming galaxy 
such as the Milky Way and the Magellanic Clouds the production of O and Ne is believed to 
be dominated by type II supernovae, we may conclude that the original O and Ne abundances 
are not significantly modified by the stellar progenitors of bright PN, which makes them 
particularly useful for chemical evolution studies. An example is the determination of 
radial abundance gradients and their variations, which provide an essential constraint 
for chemical evolution models (cf. \cite[Maciel et al. 2005]{maciel2005}, \cite[Maciel et 
al. 2006b]{maciel2006b}).

   \begin{figure}
   \centering
   \includegraphics[angle=-90,width=8cm]{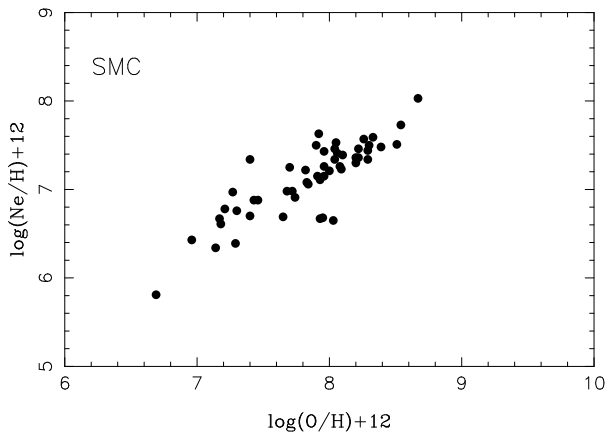}
      \caption{The Ne/H $\times$ O/H relation for the SMC.
              }
         \label{fig1}
   \end{figure}
   \begin{figure}
   \centering
   \includegraphics[angle=-90,width=8cm]{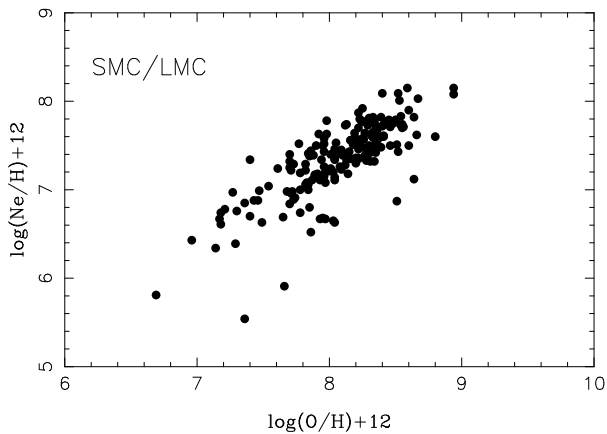}
      \caption{The Ne/H $\times$ O/H relation for the SMC and LMC.
              }
         \label{fig2}
   \end{figure}

The variation of the ratios S/H, Ar/H and Ne/H with O/H show a good positive correlation 
for all studied systems in the Local Group, with similar slopes. The main differences 
lie in the average metallicity of the different galaxies, which can be inferred from 
the observed metallicity range. The galactic nebulae extend to higher metallicities 
[up to $\epsilon$(O) = log O/H + 12 = 9.2], and the LMC objects reach $\epsilon$(O) = 8.8, 
while the lowest metallicities in the SMC are about $\epsilon$(O) = 7.0.  As an example, 
Fig.\,\ref{fig1} shows the Ne/H ratio as a function of O/H for the SMC, while Fig.\,\ref{fig2} 
includes the PN from the LMC as well. In these figures we include the combined samples by 
our group (cf. \cite[Idiart et al. 2007]{idiart}) and data from \cite[Stasi\'nska et al. 
(1998)]{stasinska} and \cite[Leisy \& Dennefeld (2006)]{leisy}.  These figures can be 
compared with Local Group galaxies from \cite[Richer \& McCall (2006)] {richer}. Other 
populations in the Galaxy also have a similar behaviour, as can be seen from the work 
of  \cite[Escudero et al. (2004)]{escudero} for PN in the  galactic bulge and 
\cite[Costa et al. (2004)]{costa} for the disk.

\begin{figure}
   \centering
   \includegraphics[angle=0,width=8cm]{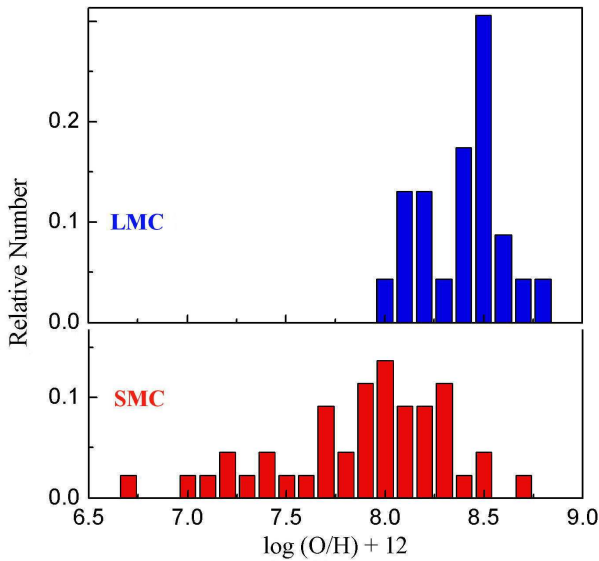}
      \caption{The metallicity distribution of PN in the Magellanic Clouds.
              }
         \label{fig3}
   \end{figure}

PN are also extremely useful to determine the metallicity distribution in a given system, 
as accurate abundances of O, S, Ne, Ar and S are often measured. A comparison of the 
distribution in different systems can be used to infer their average metallicities, with 
consequences on the star formation rates. As an example, Fig.\,\ref{fig3} shows the O/H 
distribution in the Magellanic Clouds, which can be compared with similar data 
for the galactic bulge and disk, such as given by \cite[Cuisinier et al. (2000)]{cuisinier}. 
The metallicity distribution of PN in the galactic bulge can be used as an important tool 
in order to constrain galactic chemical evolution models, as shown by 
\cite[Maciel (1999)]{maciel1999}.

\section{Chemical evolution of the SMC}

\begin{figure}
   \centering
   \includegraphics[angle=-90,width=8cm]{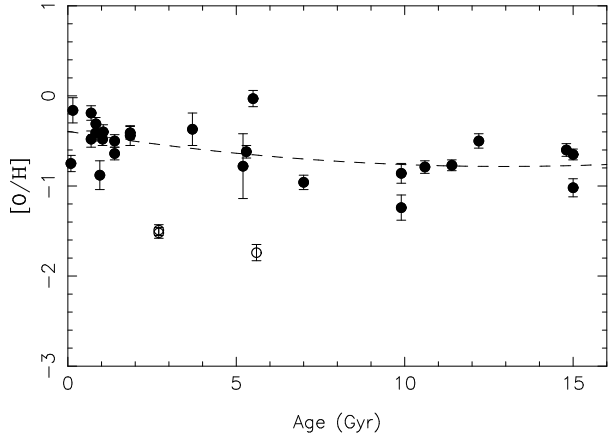}
      \caption{The age-metallicity of the SMC as derived from the O/H
      abundances. The steep increase in the metallicity near 2 Gyr
      is in good agreement with the the star formation burst as predicted
      by models by \cite[Pagel \& Tautvai\v siene (1998)]{pagel}.}
    
     \label{fig4}
   \end{figure}

We investigated the chemical evolution of the Small Magellanic Cloud  (SMC) based on 
abundance data of planetary nebulae (for details see \cite[Idiart et al. 2007]{idiart}). 
The main  goal is to study the time evolution of the oxygen abundance in this galaxy by 
deriving an age-metallicity relation, which is of fundamental importance as an observational 
constraint for chemical evolution models of  the SMC. We have used high quality PN data to 
derive the properties of the progenitor stars, so that the stellar ages could be estimated, 
using theoretical evolutionary tracks from \cite[Vassiliadis \& Wood (1992)]{vassiliadis}.
We collected  a large number of measured spectral fluxes for each nebula and derived 
accurate physical parameters (effective temperatures, luminosities, masses and ages) 
and nebular abundances. New spectral data for  a sample of PN in the SMC were obtained 
between 1999 and 2002, based on observations secured at the ESO-La Silla 1.52m and 
LNA-Brazil 1.6m telescopes. These  data are used with data available in the literature 
to improve the accuracy of the fluxes for each spectral line. We obtained accurate chemical 
abundances for about 44 PN in the SMC. Our derived oxygen-versus-age 
diagram is shown in Fig.\,\ref{fig4}. A similar relation involving the [Fe/H] metallicity was 
derived on the basis of a correlation with stellar data. We have obtained the converted  
[Fe/H] metallicities by calibrating an [O/H] $\times$ [Fe/H] relationship matching the 
abundances of the youngest objects in our sample according to the models by \cite[Pagel \& 
Tautvai\v siene (1998)]{pagel}. Taking into account the implications of the derived 
age-metallicity relation for the SMC formation, we suggest a star formation burst in the 
last 2--3 Gyr which is also apparent from Fig.\,\ref{fig4}. This behaviour is in agreement
with some recent results based on stellar clusters  and data on field stars (cf. E. Grebel, 
these proceedings), according to which there is no smooth age-metallicity relation for the SMC, 
at least during the last few Gyr.

\bigskip
{\bf Acknowledgements}. This work was partially supported by FAPESP and CNPq.

\end{document}